\documentclass{acm}
\usepackage{algorithm}
\usepackage{algorithmic}
\usepackage{amssymb}
\usepackage{amsmath}
\usepackage{subfigure}
\usepackage{graphicx}
\usepackage{enumerate}
\usepackage{verbatim}

\begin{document}

\title{Prediction Of Arrival Of Nodes In A Scale Free Network}
\numberofauthors{5} 
\author{
\alignauthor
Vijay Mahantesh SM \affaddr{intern\\Indian Statistical Institute}\\
       \affaddr{Chennai, India}\\
       \email{vijaym123@gmail.com}
\alignauthor
Sudarshan Iyengar\\ \affaddr{Indian Statistical Institute}\\
       \affaddr{Chennai, India}\\
       \email{sudarshaniisc@gmail.com}
\alignauthor Vijesh M\\
       \affaddr{Intern \\Indian Statistical Institute}\\
       \affaddr{Chennai, India}\\
       \email{mv.vijesh@gmail.com}
\and  
\alignauthor Shruthi Nayak \\
       \affaddr{Intern}\\
       \affaddr{Indian Statistical Institute}\\
       \affaddr{Chennai, India}\\
       \email{rn.shruthi@gmail.com}
\alignauthor Nikitha Shenoy\\
       \affaddr{Intern}\\
       \affaddr{Indian Statistical Institute}\\
       \affaddr{Chennai, India}\\
       \email{nikithashenoyk@gmail.com}
}
\maketitle
\begin{abstract}
\hspace{0.17in} Most of the networks observed in real life obey power-law degree distribution. It is hypothesized that the emergence of such a degree distribution is due to preferential attachment of the nodes. Barabasi-Albert model is a generative procedure that uses preferential attachment based on degree and one can use this model to generate networks with power-law degree distribution. In this model, the network is assumed to grow one node every time step. After the evolution of such a network, it is impossible for one to predict the exact order of node arrivals. We present in this article, a novel strategy to partially predict the order of node arrivals in such an evolved network. We show that our proposed method outperforms other centrality measure based approaches. We bin the nodes and predict the order of node arrivals between the bins with an accuracy of above $80\%$. 
\end{abstract}


\section{Introduction}

\hspace{0.17in} Real world networks such as biological, social and technological networks  are the products of an evolutionary process. These networks are generally classified as Scale Free Networks (SFN) by nature. SFNs are a class of networks in which degree distribution follows Power Law. Generative models such as Duplicate-Mutation, Forest Fire and Preferential Attachment ~\cite{barabasi02} have been proposed to synthesize SFNs. The synthesis of dynamic SFNs involves a continuous addition of new nodes to the existing network. The behavior of each new node depends on the generative model being used. It is interesting to study how nodes get assembled in complex network over time ~\cite{sanket11}. Given the snapshot of a dynamic network, is it possible to probabilistically predict the evolutionary sequence of the nodes in the network? \\

We propose a method that predicts the order of arrival of nodes in the given Scale-Free Network, modeled and synthesized using a specified generative model. This approach first computes a vertex ranking of the given network based on a ranking methodology. We then synthesize several such networks using the generative model that was used in the construction of the given network. It is important to note that the order of arrival of nodes in the synthesized networks is known. The same ranking methodology is applied to compute the vertex ranking for each of the synthesized networks. The nodes in the given network are mapped to the nodes in a synthesized network, according to a bijection function between the vertex rankings. We then predict the probable order of arrival of nodes in the given network, based on the bijective mapping and the order of arrival of nodes in the synthesized network. This method of mapping, over several such synthesized networks,  associates a probability with every pair of vertices. This probability denotes the arrival order of vertices in the corresponding vertex pair.

We then construct a Directed Graph (DG) by drawing an edge for every pair in their predicted order of arrival. We propose a binning methodology, wherein the nodes of the DG having similar characteristics are grouped into hypothetical containers called bins. The order of arrival of nodes within a bin is unknown. Hence, we determine the order of arrival of nodes across several such bins.

\section{Preliminaries and Notations}
\subsection{Scale Free Networks}
\hspace{.18in} A Scale-Free Network (SFN) is a network whose degree distribution follows a \emph{power law}. 
Many real world networks are known to exhibit a decaying degree distribution.
This kind of distribution is called a power law. Mathematically, it is defined as
\begin{equation}
\fbox{$\bf P(k) \approx ck ^{-\gamma}$} 
\end{equation}
where, \\
\textbf{k} is \emph{degree},\\ 
\hspace{.18in}\textbf{c}  is a \emph{normalization constant} and \\
\textbf{$\gamma$} is a \emph{parameter} whose value is typically in the range (2,3)\\

The high degree nodes in a SFN are often called as "hubs". 
The power law degree distribution of the SFNs suggests the existence of a small number of high degree nodes. Although the hubs are small in number, they dominate the network to a great extent. 
Removal of the hubs from the network might cause a network breakdown and disrupt the network characteristics. Figure 1 shows an example of a SFN. The degree distribution of the same network is shown in Figure 2.
\begin{figure}[htp]
\centering
\includegraphics[scale=0.08]{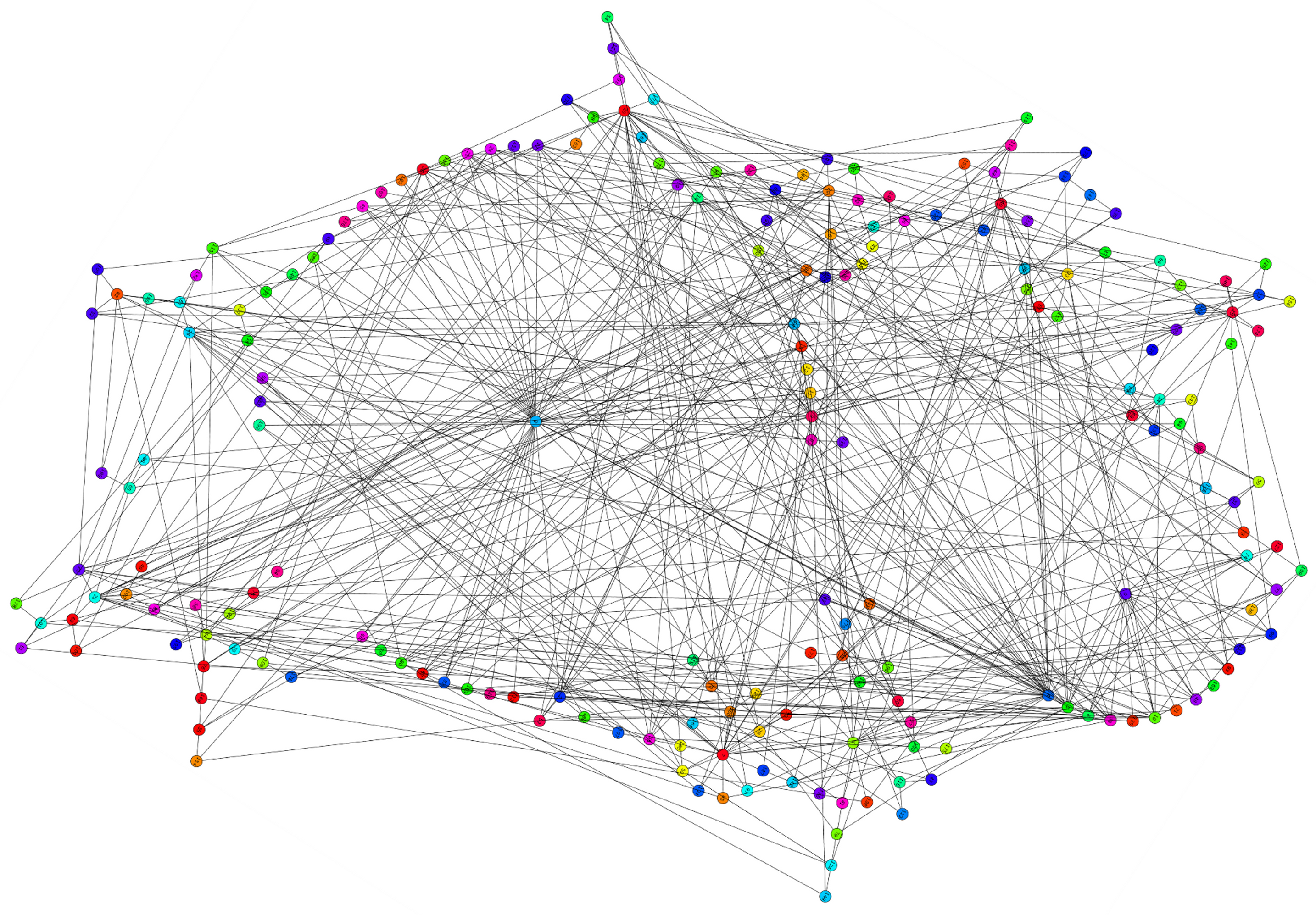}
\caption{A Scale-Free Network of 200 nodes.}
\label{}
\end{figure}
\begin{figure}[htp]
\centering
\includegraphics[scale=0.16]{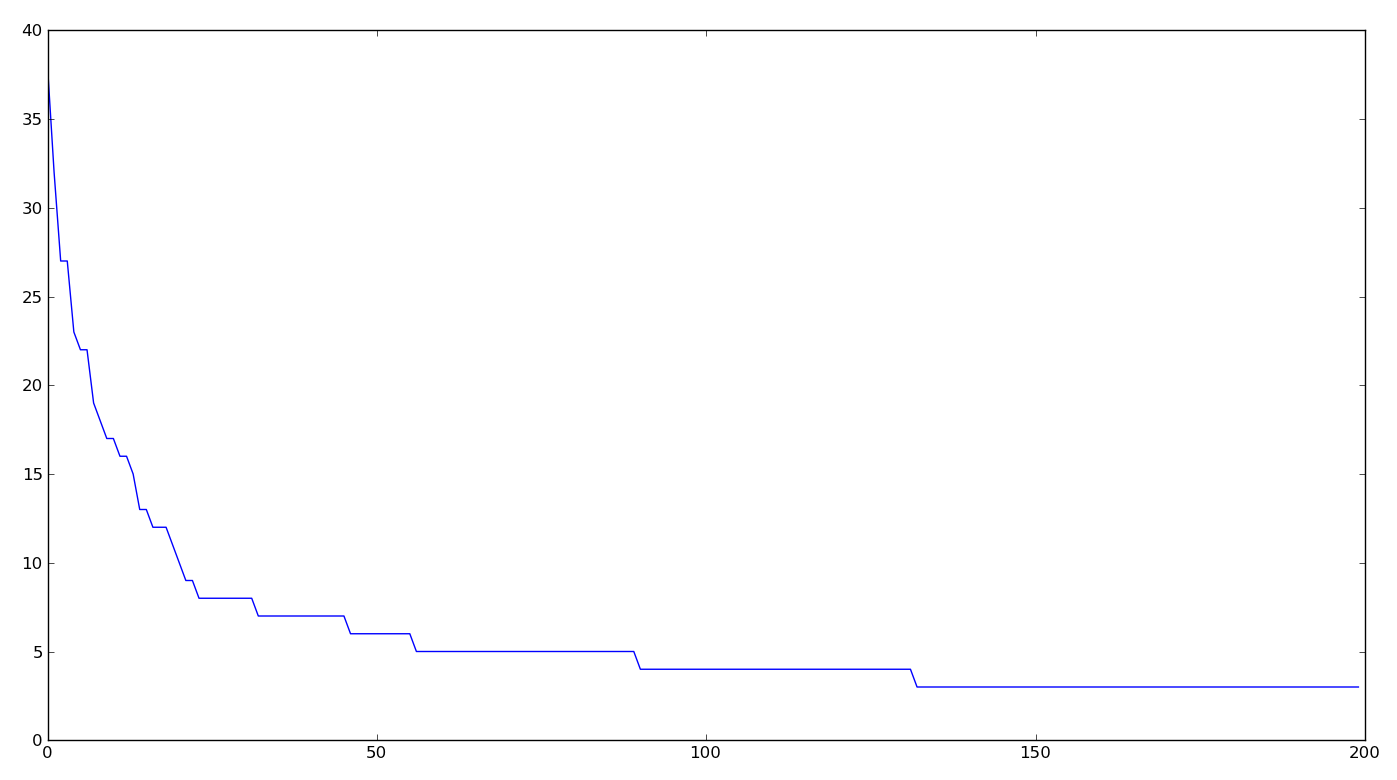}
\caption{The degree distribution curve for the network in Figure 1. This network follows a power law degree distribution.}
\label{}
\end{figure}

\subsubsection{Generative Model for Scale Free Networks}
\hspace{.18in} To explain the power law degree distribution in the real world networks, mechanisms such as preferential attachment and fitness model, etc..have been proposed. 
Barabasi and Albert proposed a randomized algorithm for generating SFNs using a preferential attachment mechanism. 
This model is referred to as BA model~\cite{barabasi99-1}.

\newpage
\textbf{Algorithm to construct a BA Network $G(V_{final},C)$ :}

Let $C$ be the number of connections that each new node must create on its arrival.
Let $V_{final}$ be the vertex set of the completely generated network G. It is clear that $|V_{final}| > C$.
As the network evolves, let $V$ and $E$ be the instantaneous vertex set and edge set of the intermediate networks respectively.

\begin{algorithmic}
\STATE The nodes are designated by enumerating them as 
\STATE $\{0,1,2, . . . , (|V_{final}|-1)\}$.
\STATE A complete network $K_C$ with $C$ nodes is constructed. Now, $|V|=C$.
\WHILE {$|V| \le |V_{final}|$}
  \STATE Generate a new node $u$.
  \STATE Preferential Attachment: Let $v \in V$ be sampled according to the cumulative degree distribution function, $CDF(i)$.
	   \begin{equation}
	    CDF(i) = \sum^{i}_{j}\frac{degree(Nj)}{2*total\_edges}\ \ where\ N_j \in V
	   \end{equation}
  \STATE $iter \leftarrow 1$
  \WHILE {$iter \le C$}
    \STATE  Let $r$ be a real number uniformly picked at random in [0,1).
    \STATE  Choose $u \in V\ |\ CDF(u-1) \le r < CDF(u)$.
    \IF {$(u,v) \notin E$}
      \STATE append $(u,v)$ to $E$
    \ELSE
      \STATE $iter \leftarrow iter - 1$
    \ENDIF
    \STATE $iter \leftarrow iter + 1$
  \ENDWHILE
\ENDWHILE
\end{algorithmic}

Figure 3 illustrates the growth of a BA Network $G(9,3)$.
\begin{figure}[htp]
\centering
\includegraphics[scale=0.25]{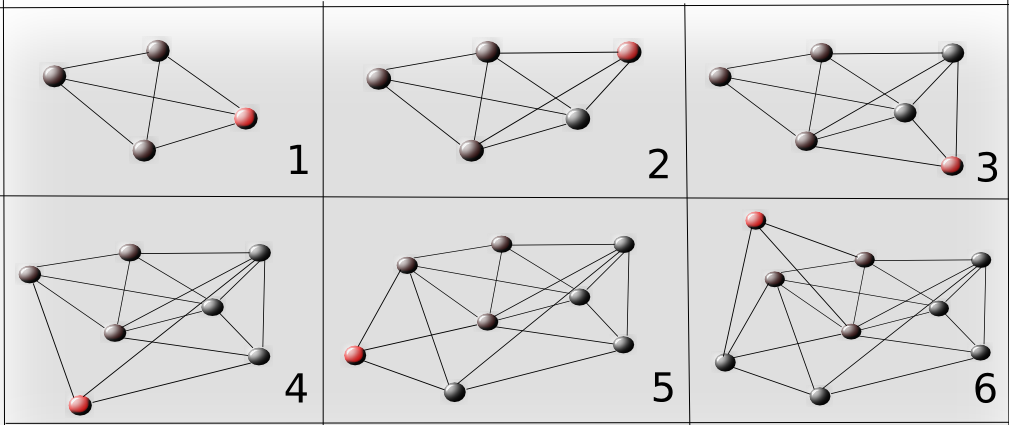}
\caption{Growth of a BA Network with 9 nodes and 3 connections.}
\label{}
\end{figure}

\newpage
\subsection{Directed Acyclic Graph}
\hspace{.18in} A Directed Acyclic Graph (DAG) is a directed graph containing no cycles. 
Indegree of a node $v$ in a directed graph $G$ is defined as $|S| : S \leftarrow \{(u,v) | (u,v) \in E_G\}$. 
It is denoted by $InDegree(v)$.
Outdegree of a node $v$ in a directed graph is defined as $|S| : S \leftarrow \{(v,u) | (v,u) \in E_G\}$.
It is denoted by $OutDegree(v)$.

\subsection{Lists and Index of an element}
\hspace{.18in} A list is an ordered set of elements. 
Index of an element $u$ in a list $L$ is the position at which the element $u$ occurs in $L$, denoted by $index_{L}(u)$.

\subsection{Centrality Measures}
\hspace{.18in} A centrality measure is a function that associates a real value with each vertex in a network~\cite{borgatti06}.
The value indicates how central or important the vertex is, in the network. 
Here, the term ``important'' is application specific.
This gives rise to many centrality measures, each of which rates the nodes according to some property of the node.

\subsubsection{Degree Centrality}
\hspace{.18in}  Degree of a node is often interpreted as an effective measure of influence or importance of that node in a network. Degree of a node $u$ in a graph in denoted by $deg(u)$~\cite{bavelas50}.
The Degree Centrality assigns a node $u$ with a value that is proportional to $deg(u)$.
\\
Mathematically, for a graph $G(V,E)$:
\begin{equation}
C_{degree}(v) = \frac{deg(v)}{|V|-1} 
\hspace{.4in} v \in V
\end{equation} 

\subsubsection{Betweenness Centrality}
\hspace{.18in}  Betweenness Centrality assigns a node $v$ with a value that is proportional to the number of shortest paths ~\cite{anthonisse71} ~\cite{freeman77}, between all other pairs of vertices, that pass through $v$.

Let \rm$\delta(v)$ denote the fraction of shortest paths between $s$ and $t$ that contain the vertex $v$:
\begin{equation}
\delta_{st}(v) = \frac{\sigma_{st}(v)}{\sigma_{st}} 
\end{equation}
\hspace{0.17in} where $\sigma_{st}$ denotes number of all shortest paths from vertex $s$ to $t$ and $\sigma_{st}(v)$ denotes the number of shortest paths from $s$ to $t$ passing through $v$.
Then the Betweenness Centrality of a vertex $v$ is given by 
\begin{equation}
C_{betweenness}(v) = \sum_{s\neq v \neq t \in G} \delta_{st}(v)
\end{equation}
In our experiments, we have used Brandes approach to compute betweenness centrality ~\cite{brandes01}.
\subsubsection{Eigenvector Centrality}
\hspace{.18in} The index in Eigenvector Centrality characterizes the individuals in connected networks according to their level of popularity ~\cite{bonacich72} ~\cite{zweig05}.
It is a more sophisticated version of Degree Centrality. 
A given node is said to be popular if it is connected to many other nodes or few nodes with a very high popularity.
Mathematically, this can be formulated as follows:

Let $A$ be the adjacency matrix of the network $G(V,E)$. $A_{u,v} = 1$ if $(u,v) \in E_G$ and $A_{u,v} = 0$ if $(u,v) \notin E_G$.
Let $x_u$ denote the centrality score of $u \in V_G$. 
$x_u$ is proportional to the sum of the scores of $neighbors(u)$. Hence
\begin{equation}
x_{u} = \frac{1}{\lambda} \sum_{v=1}^{|V|} A_{u,v}\ x_v
\end{equation}
where $\lambda$ is a constant.\\

On defining $x = [x_0\ x_1\ x_2\ ...\ x_{|V|-1}]$ as a vector of centrality scores, we can transform the above equation into a matrix form as 
\begin{equation}
x = \frac{1}{\lambda} A x
\end{equation} 
Assuming that we wish the centrality scores to be a non-negative real value, it can be shown (using the Perron-Frobenius theorem) that $\lambda$ must be the largest Eigen Value of $A$. 
$x$ is the Eigen Vector corresponding to the Eigen Value $\lambda$.
\\

\subsection{Reference Network}
\hspace{.18in} In our experiments, we study the SFNs generated using the Barabasi-Albert Model.
Let $G_m(V_m, C_m)$ represent a Barabasi-Albert Network whose vertex arrival order is to be deduced.
For evaluative purposes, we record the order of arrival of vertices in $G_m$ during its inception. 
Let $list_{true}$ be a sequence of vertices that represent the actual order of arrival of vertices in $G_m$.
We will be referring to $G_m(V_m, C_m)$ in all the further sections as the input network to the proposed algorithm that predicts order of arrival of nodes.

\section{Centrality Measure based \\ Methods}
\subsection{Degree Binning}
\hspace{.18in} The degree of a node is the number of connections associated with that node. 
A naive approach towards the solution to the vertex arrival order prediction problem is to exploit and explore the contribution of this factor.

From the preferential model of SFN construction, it is evident that the last few nodes that get connected to the network will have a relatively low degree, as compared to the nodes that had arrived in the initial stages. 
Consider the network $G_m$ from section 2.5.
Intuitively, we hypothesize that higher the degree of a node, higher is its influence in the network, and earlier it has arrived during the network evolution.
We can state with a high probability, that the notable hubs in $G_m$ would have arrived prior to the nodes with a relatively low degree.

Hence, we rank the nodes in the decreasing order of their degree.
The equi-degree nodes are assigned with the same ranking. 
We then place the vertices with the same ranking into a hypothetical container, referred to as a bin.
The ranking of a bin is same as the ranking of node(s) inside the bin. 
The number of bins formed is the total number of unique ranks assigned to the nodes. 
We then apply a Binning Quality Measure (BQM) to compute the accuracy of our prediction of order of arrival of nodes \emph{across} the bins. 
BQM quantifies the prediction accuracy on a scale of 0 to 1.
Figure 5 illustrates the Binning Methodology that we use to predict the order of arrival of nodes across the bins.

\begin{figure}[htp]
\centering
\includegraphics[scale=0.25]{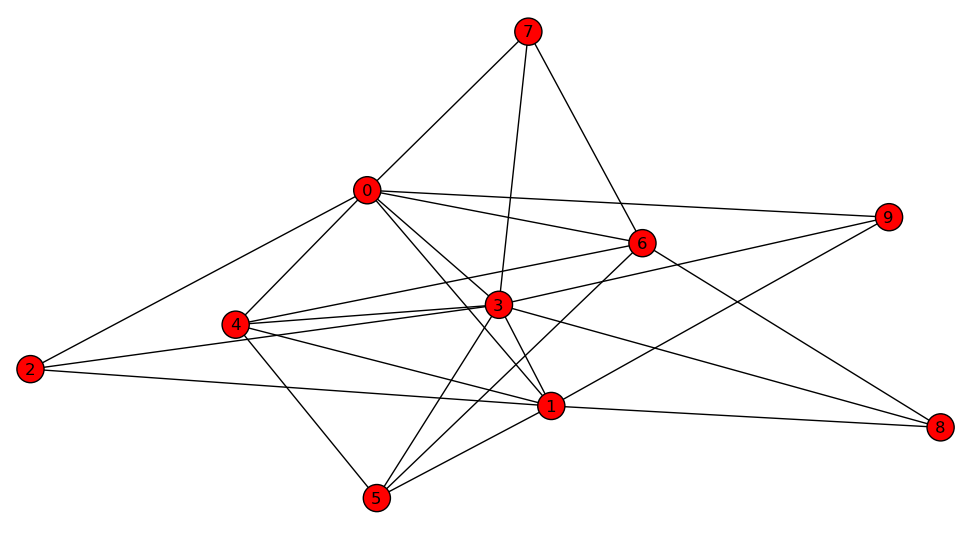}
\caption{A SFN, constructed using BA model with 9 nodes and 3 connections.}
\label{}
\end{figure}

\begin{figure}[htp]
\centering
\includegraphics[scale=0.23]{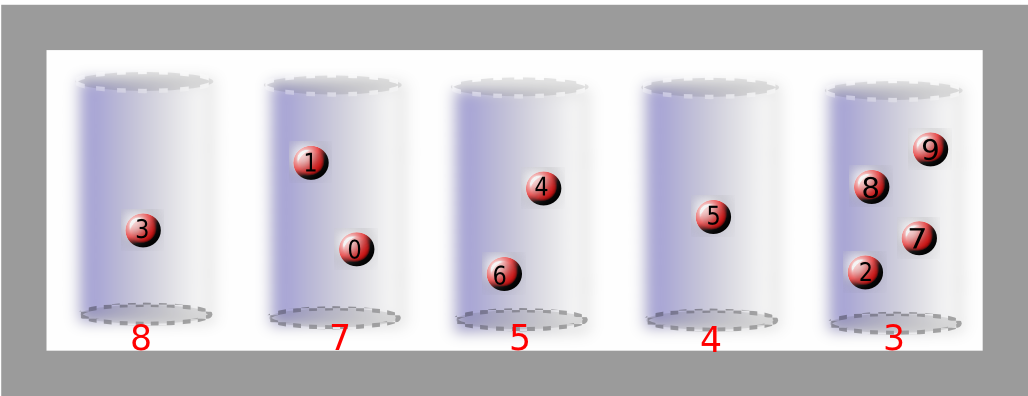}
\caption{Binning the nodes of the network in figure 4 based on degree. The numbers below the bins denote the degree of the nodes that are present in the bin.}
\label{}
\end{figure}

The following mathematical formulation illustrates a technique to quantify the correctness of our prediction.
We refer to the technique as Binning Quality Measure (BQM). \\
Let $\delta$ be the number of bins. 
Let $B = [B_0, B_1, B_2, ... ,B_{\delta}]$ be the predicted chronological bin ordering.
We associate a score $\beta$ between every pair of bins. The final prediction measure $\eta$ is computed as a ratio of sum of $\beta$ for all bin-pairs and the total number of bin-pairs.\\

To calculate $\beta$ for a pair of bins $B_i$ and $B_j$, with $i<j$:\\
Here, we claim that the nodes in $B_i$ has arrived before the nodes in $B_j$
Hence, we impose the condition $i<j$, with reference to the predicted chronological bin ordering $B$.\\
For a pair of vertices $u \in B_i$ and $v \in B_j$, we define\\

$vertexOrder(u,v) = 1\ if\ index_{list_{true}}(u) < index_{list_{true}}(v)$ \\

$vertexOrder(u,v) = 0\ if\ index_{list_{true}}(u) > index_{list_{true}}(v)$\\

$\beta(i,j) = \frac{ \sum_{u \in B_i, v \in B_j}vertexOrder(u,v) }{ |B_i| |B_j| }$\\

The final prediction measure $\eta$ is given by
\begin{equation}
\eta = \frac{ \sum_{0 < i < j \le \delta}{\beta(i,j)} }{ ^{\delta}C_{2} }
\end{equation}

\subsection{Binning based on Centrality Meaures}
\hspace{.2in} The main drawback of binning based on degree is that, the degree centrality indices associated with the nodes are not distinct in $G_m$. This is because there can exist many number of nodes with the same degree. Hence, binning based on degree centrality results in a small number of bins, with a large number of nodes per bin. Ideally, it is desirable to have more number of bins with a less number of nodes per bin.

We move on to yet another approach which could provide us with a large number of bins. In this approach, we apply $\chi$ centrality to main graph.  Based on an intuitive conjecture, higher the $\chi$ centrality a node, earlier it has arrived in the network evolution. Hence, we sort the vertices in the decreasing order of their $\chi$ centrality indices. We group the nodes from this sorted ordering into $\delta$ number of bins, each bin containing $\frac{|V_{G_m}|}{\delta}$ number of nodes. We refer to the list of bins thus obtained as $binOrdering_\chi$. In our experiments, we choose $\chi$ to be Betweenness Centrality and Eigenvector Centrality. We use BQM (refer section 3.1) to quantify the accuracy of the prediction using binning based on centrality.
\section{A New Vertex Ranking:\\ Differential Core Ranking}
\hspace{0.18in} In this section, we formulate a new method of ranking nodes. Let $G(V,E)$ be any graph. Let $DCR_G$ represent the Differential Core Ranking of G.

Let $\chi$ be any centrality measure. Let $G_0$ be the initial graph. Let $G_1$ be the graph obtained from $G_0$ after removal of nodes with the minimum degree. The change in $\chi$ centrality value of the nodes in $G_0$ is set as the attribute of the corresponding node. We then apply the above procedure starting with $G_1$. Let $G_2$ be the graph obtained from $G_1$ after the removal of nodes with the minimum degree. The change in the $\chi$ centrality value of the nodes in $G_1$ is added to the attribute of the corresponding node.\\
In general, let $G_{i+1}$ be the graph obtained from $G_i$ after the removal of nodes with the minimum degree. The change in the $\chi$ centrality value of the nodes in $G_i$ is added to the attribute of the corresponding node. This procedure is repeated until there are no nodes left in $G_i$.

The algorithm to compute $DCR_G$ is as follows:
\begin{algorithmic}
\STATE Let $\chi$ represent any centrality measures
\STATE Let $G_0$ represent the given graph $G$
\STATE Let $u \in V(G)$. Let the Differential Core Measure $DCM_u$ be a value associated with $u$.\\
Set $DCM_u = 0\ \forall u \in V(G)$
\STATE Let $\chi_{u,G_k}$ represent the $\chi$ centrality value of $u$.
\STATE Let $i \leftarrow 0$
\WHILE{$|V_{G_i}| > 0$}
\STATE Let $minDeg \leftarrow arg min(deg(u)), u \in V(G_i)$
\STATE Let $minVertices \leftarrow \{ u_0, u_1.... u_n \}, deg(u_m) = minDeg$
\STATE Let $G_{i+1} \leftarrow graph obtained\ after\ removing$\\ 
 \hspace{2.0in}$minVertices\ from\ G_i$
\STATE $DCM_u \leftarrow DCM_u + abs(\chi_{u,G_{i+1}} - \chi_{u,G_{i}})\ \forall u \in V(G_{i})$\\
\hspace{2.0in}$and\ u \in V(G_{i+1})$
\STATE $DCM_u \leftarrow DCM_u + abs(0 - \chi_{u,G_{i}})\ \forall u \in V(G_{i})\ and \ u \notin V(G_{i+1})$
\STATE $i \leftarrow i + 1$
\ENDWHILE
\STATE $DCR_G \leftarrow \{(DCM_{u_0},u_0), (DCM_{u_1},u_1)...,(DCM_{u_{|V_G|}},u_{|V_G|})\}$
\end{algorithmic}
$DCR_G$ gives the Differential Core Ranking of the vertices. $DCM_u$ denotes the centrality score of the node $u$. Higher the sum of changes in the $\chi$ centrality values of a node, higher is its importance in the network.

\section{Network Reconstruction \\ Algorithm}
\hspace{0.17in} In this section of the paper, we describe our algorithm to predict the order of arrival of nodes in $G_m$.\\

Our Algorithm is mainly divided into 4 subsections. Section 5.1 aims at generation of Synthetic Networks that resemble $G_m$. Section 5.2 describes a mapping procedure and derivation of prediction lists. In section 5.3, we analyze the prediction list and construct a directed graph. Section 5.4 deals with the transformation of directed graph to a directed acyclic graph and binning of nodes.

\subsection{Generation of Synthetic Networks}
\hspace{0.17in} The main focus of this section of the algorithm is to recreate the growth environment of the reference network $G_m$. Since the exact replication of $G_m$ is not possible, we generate networks that are similar to $G_m$ in certain characteristics. We refer to these set of networks as Synthetic Networks.

Let $\alpha$ be the number of Synthetic Networks generated. Let $S_i$ and $chronology_{S_i}$ denote the Synthetic Network and the order of arrival of nodes in the corresponding $S_i$. In our experiments, we use BA model to generate $S_i$, with $|V_m|$ number of nodes and $C_m$ connections. It is worth noting that every time we generate a Synthetic Network $S_i$, we keep track of the network growth by recording $chronology_{S_i}$. Since the Synthetic Networks are built on the same model as that of $G_m$, we hypothesize that the chronology of $S_i$ is \emph{similar} to the actual order of arrival of nodes in $G_m$. Hence, it is righteous to make use of $chronology_{S_i}$ in predicting the probable order of arrival of nodes in $G_m$.

\subsection{Mapping and Derivation of Prediction Lists}

\hspace{0.17in} We have now generated $\alpha$ number of BA Synthetic Networks that is similar to $G_m$ in terms of the number of vertices $|V_m|$ and connections $C_m$. 
The chronology of the Synthetic Networks $S_i$, where $1 \le i \le \alpha$, is known. 
In this section, we intend to derive an ordering of nodes in $V_m$, corresponding to each $S_i$. 
This ordering of nodes is the predicted order of arrival of nodes in $G_m$ (during its inception), derived in accordance with $chronology_{S_i}$.
We refer the node ordering corresponding to $S_i$ as $PredList_i$. 
The procedure that we follow to deduce $PredList_i$ is explained in the remainder of the section.

We apply DCR, with $\chi$ as the base centrality measure (Refer to section 2.4), to $G_m$ in order to obtain $DCR_{G_m}$. 
$DCR_{G_m}$ is a list of vertex rankings sorted according to their DCM values. (Refer to section 4)

Consider a Synthetic Network $S_i$. 
We apply DCR, with $\chi$ Centrality as the base centrality measure, to $S_i$ in order to obtain $DCR_{S_i}$. 

Both $DCR_{G_m}$ and $DCR_{S_i}$ lists the vertices of $G_m$ and $S_i$ respectively in the decreasing of their importance. 
Lower the position of a vertex in these lists, higher its importance in the corresponding network. 
A direct bijection mapping is carried out  between $DCR_{G_m}$ and $DCR_{S_i}$. 
This mapping maps the equi-important vertices in both the networks.

Mathematically, we define a mapping function as:\\
Let $f_{map}:V_{S_i}\rightarrow V_{G_m}$ be a direct bijection between $V_{S_i}$ and $V_{G_m}$\\
i.e, $f_{map}(u) = v$ where $u \in V_{S_i}, v \in V_{G_m}$ and $index_M(u) = index_N(v)$

We propose that the nodes of equal importance in $G_m$ and $S_i$ have the same chronological ranking. 
Since we know $chronology_{S_i}$, we deduce $PredList_{S_i}$ by replacing each vertex $u$ in $chronology_{S_i}$ with $f_{map}(u)$.

We repeat the above procedure for each $S_i$.
At this stage, we have $\alpha$ prediction lists, denoted by $PredList_i$, each corresponding to a particular $S_i$.

Algorithm for Mapping:

\begin{algorithmic}
\STATE Input: The Reference Network $G_m$ and Synthetic Networks $\{S_1,S_2, ... S_{\alpha}\}$
\STATE Output: $\alpha$ Prediction Lists
\STATE Apply DCR, with $\chi$ as the base centrality measure, to $G_m$ 
\STATE Let $u_i \in V_m : 1 \le i \le |V_m|$
\STATE Let $DCR_{G_m}(u_i)$ denote the DCR associated with the vertex $u_i$
\STATE Let the tuple list $M \leftarrow \{(DCR_{G_m}(u_1),u_1),(DCR_{G_m}(u_2),u_2),$ \\
\hspace{1.8in}$... (DCR_{G_m}(u_{|V_m|}),u_{|V_m|})\}$ 
\STATE Sort $M$ in the descending order of $DCR_{G_m}(u_i)$
\FORALL {$i = 1\ to\ \alpha$}
  \STATE Let $v_j \in V_{S_i} : 1 \le j \le |V_{S_i}|$
  \STATE Let $(v_1,v_2, ... v_{|V_{S_i}|})$ denote $chronology_{S_i}$
  \STATE Apply DCR, with $\chi$ centrality as the base centrality, to the Synthetic Network $S_i$
  \STATE Let $DCR_{S_i}(v_j)$ denote the DCR of the vertex $v_j$
  \STATE Let the tuple list $N \leftarrow \{(DCR_{S_i}(v_1),v_1),(DCR_{S_i}(v_2),v_2),$\\
\hspace{1.8in} $... (DCR_{S_i}(v_{|V_{S_i}|}),v_{|V_{S_i}|})\}$
  \STATE Sort $N$ in the descending order of $DCR_{S_i}(v_j)$
  \STATE Let $f_{map}:V_{S_i}\rightarrow V_{G_m}$ be a bijection between $V_{S_i}$ and $V_{G_m}$
  \STATE $f_{map}(u) = v$ where $u \in V_{S_i}, v \in V_{G_m}$ and $index_M(u) = index_N(v)$
  \STATE $PredList_i \leftarrow (f_{map}(v_1),f_{map}(v_2), ... f_{map}(v_{|V_{S_i}|}))$.
\ENDFOR
\end{algorithmic}

Figures [6 to 9] illustrate an instance of Mapping of nodes between $G_m$ and any $S_i : 1 \le i \le \alpha$.
Figure 10 illustrates the derivation of prediction list $PredList_i$ using $chronology_{S_i}$.
   
\begin{figure}[htp]
\centering
\includegraphics[scale=0.10]{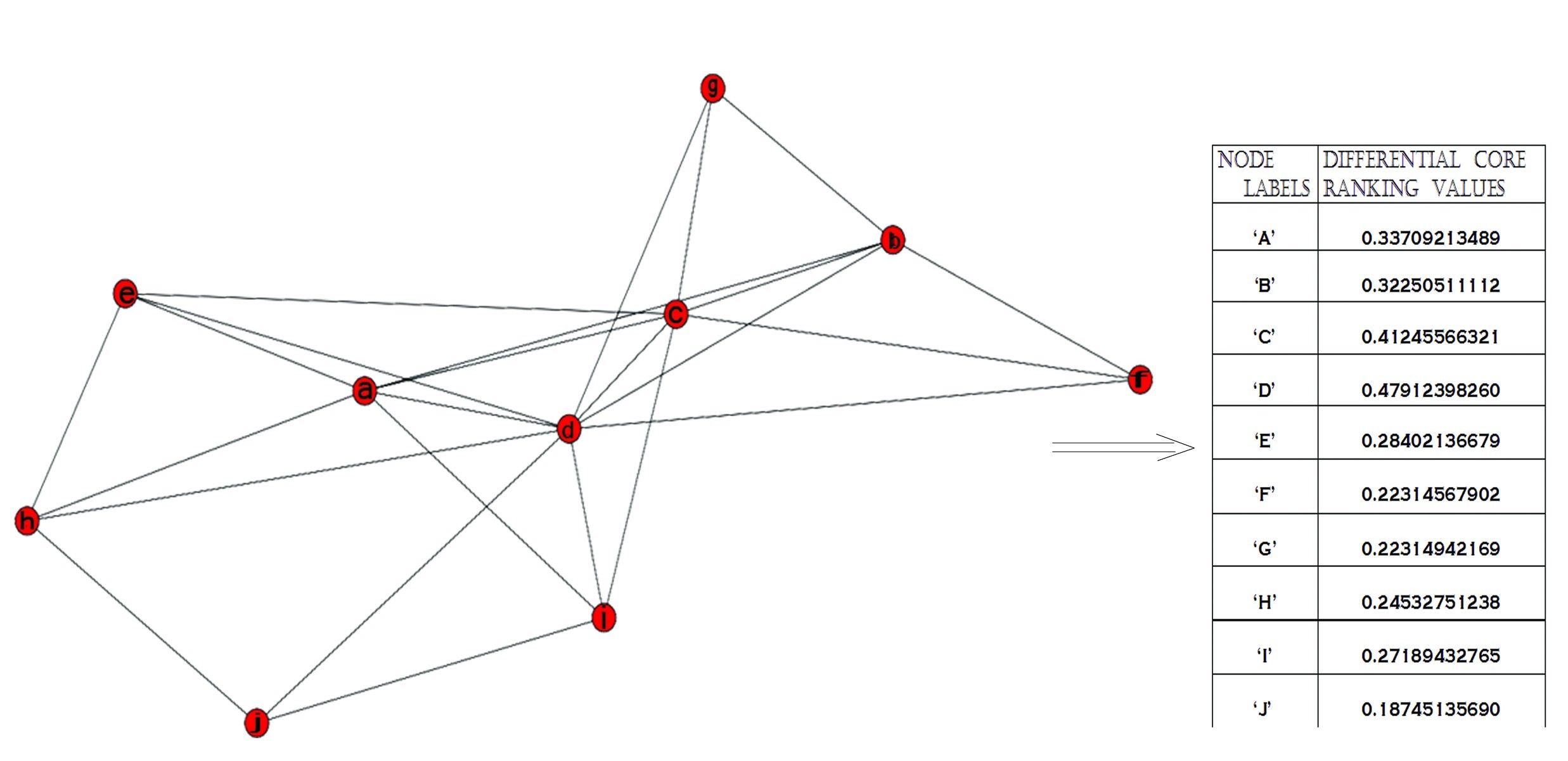}
\caption{Applying Differential Core Ranking, with Betweenness Centrality as the base centrality, to $G_m$.}
\label{}
\end{figure}

\begin{figure}[htp]
\centering
\includegraphics[scale=0.10]{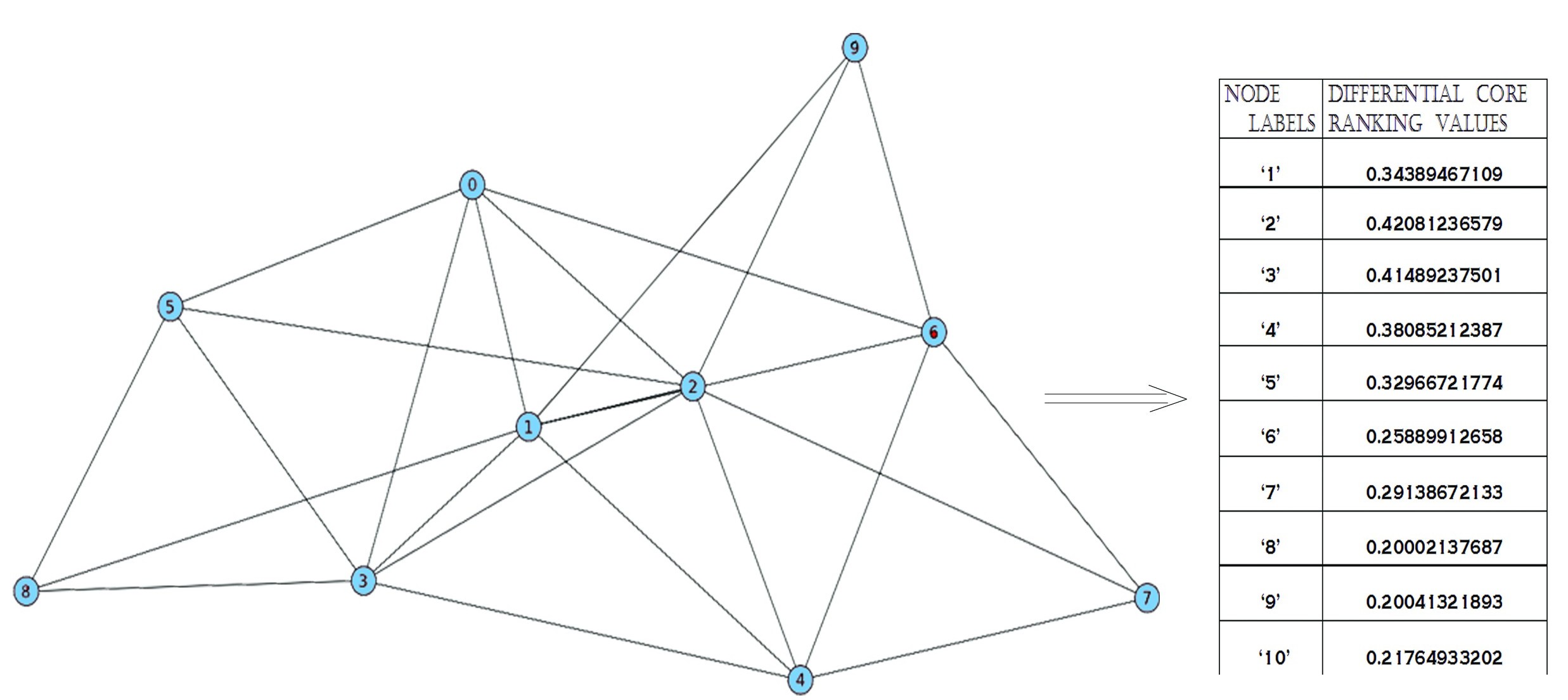}
\caption{Applying Differential Core Ranking, with Betweenness Centrality as the base centrality, to one of the $S_i : 1 \le i  \le \alpha$.}
\label{}
\end{figure}

\begin{figure}[htp]
\centering
\includegraphics[scale=0.10]{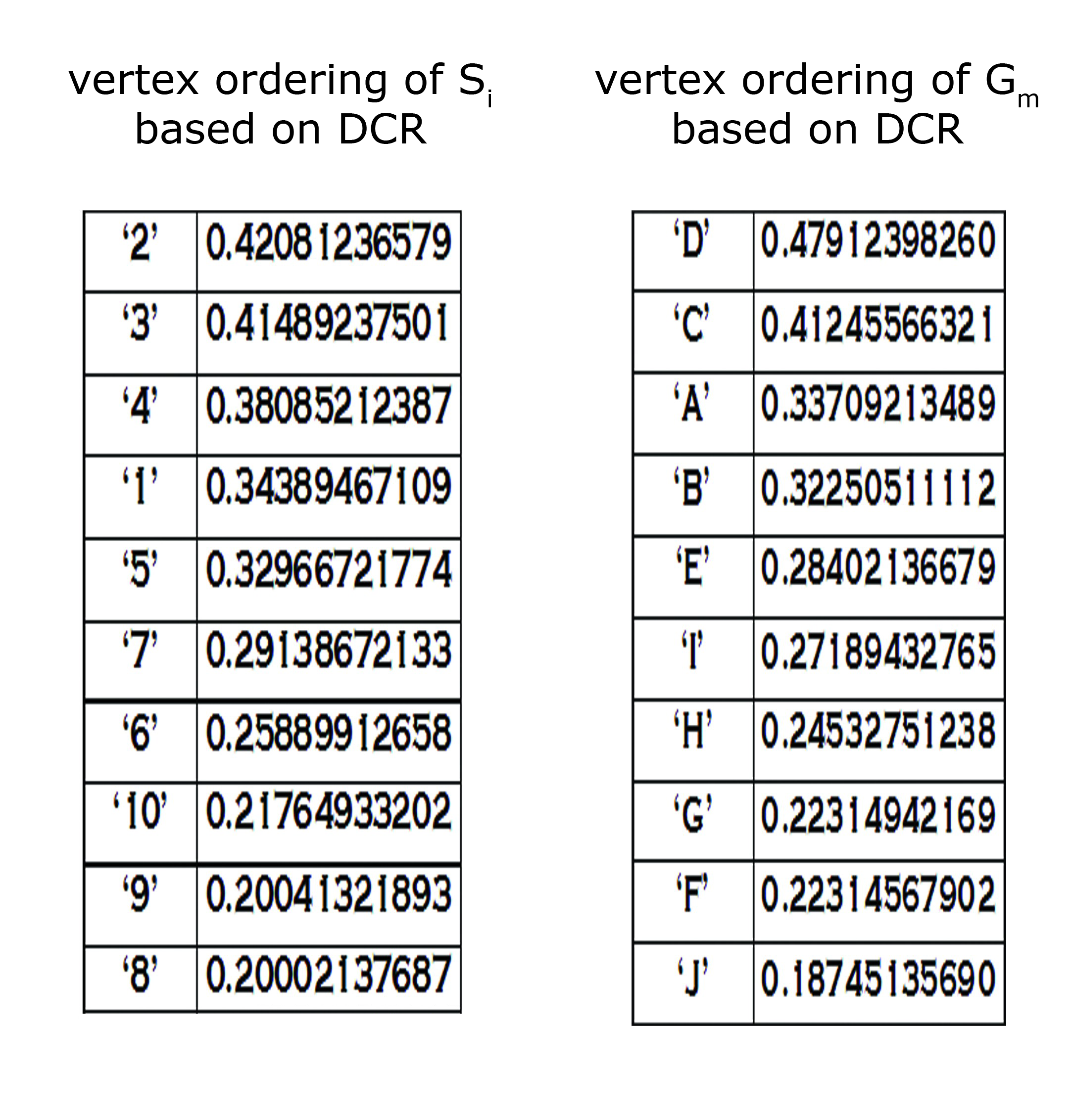}
\caption{Vertex ordering based on decreasing Differential Core Ranking for $V_{G_m}$ and $V_{S_i}$.}
\label{}
\end{figure}

\begin{figure}[htp]
\centering
\includegraphics[scale=0.08]{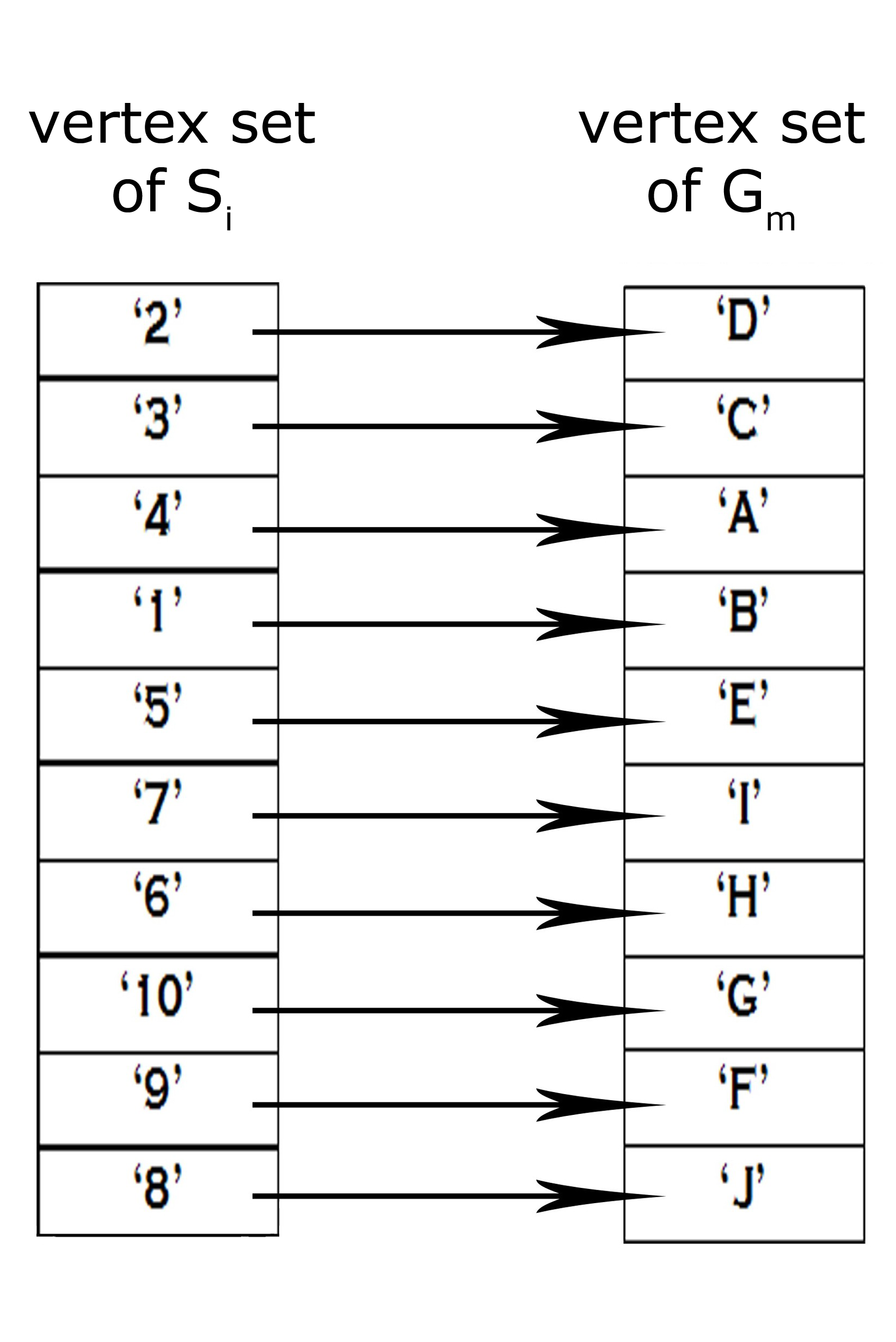}
\caption{Direct bijection mapping of vertices between Lists in figure 8.}
\label{}
\end{figure}

\begin{figure}[htp]
\centering
\includegraphics[scale=0.08]{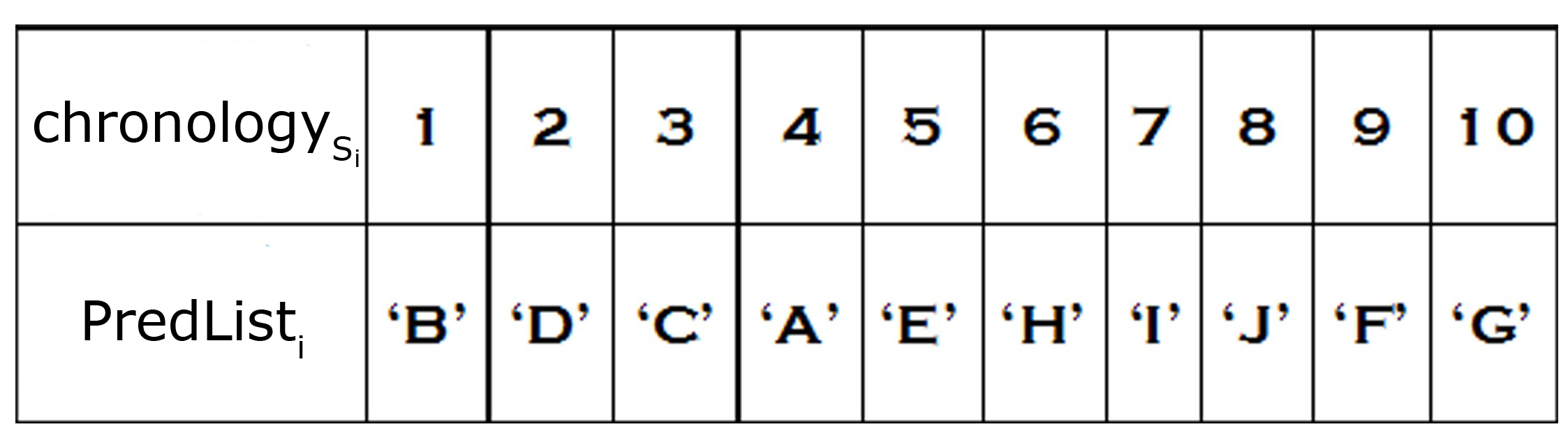}
\caption{Deduction of $PredList_i$ by reordering the nodes of $V_m$ according to $chronology_{S_i}$.}
\label{}
\end{figure}

\newpage
\subsection{Analysis of Prediction Lists and Construction of Directed Graph}
\hspace{0.17in} In the previous section, we have deduced $\alpha$ number of Prediction Lists,  $PredList_i : 1 \le i \le \alpha$. For every pair of vertices $(u,v) : u,v \in V_{G_m}$, we find the order of occurrence of $u$ and $v$ in each $PredList_i$. Let $P_{(u,v)}$ denote the probability of $u$ arriving before $v$ during the inception of $G_m$. We compute $P_{(u,v)}$ as the fraction of the number of times $u$ has occurred before $v$ in the $\alpha$ Prediction Lists. By intuitive reasoning, it is not hard to infer that, if $P_{(u,v)} < 0.5$, then $v$ has probably arrived before $u$ during the inception of $G_m$. Hence, we set $P_{(v,u)} = 1 - P_{(u,v)}$. We then construct a Directed Graph $DG$ with vertex set $V_{DG} = V_m$, and edge set $E_{DG}=\phi$. A directed edge from $u$ to $v$ in $DG$ indicates that $u$ has arrived before $v$ during the construction of $G_m$. For a pair of vertices $(u,v)$:\\
  if $P_{(u,v)} > 0.5$, then we say that $u$ has arrived before $v$ with a probability $P_{(u,v)}$\\
  if $P_{(u,v)} < 0.5$, then we say that $v$ has arrived before $u$ with a probability $1 - P_{(u,v)}$

\newpage
The algorithm to deduce $DG$ is presented below:\\
\begin{algorithmic}
  \STATE Let $S \leftarrow \{S_1,S_2,S_3, ... S_{\alpha}\}$ denote the set of Synthetic Networks
  \STATE Let $PredList_i$ denote the Prediction List corresponding to $S_i : 1 \le i \le \alpha$ (Refer to algorithm in section 5.2)
  \STATE Construct a Directed Graph $DG_m$ with $V_{DG_m} = V_m$ and $E_{DG_m} = \phi$
  \STATE Let $P_{(u,v)}$ be the probability associated with $(u,v) : u,v \in V_{DG_m}$ in determining if $u$ has come before $v$.
  \FOR {all unordered pairs $(u,v) : u,v \in V_m\ and\ u \ne v$}
    \STATE $count \leftarrow 0$
    \FOR {$i \leftarrow 1\ to\ \alpha$}
      \IF {$index_{S_i}(u) < index_{S_i}(v)$}
	\STATE $count \leftarrow count + 1$
      \ENDIF
    \ENDFOR
    \STATE $P_{(u,v)} \leftarrow count/\alpha$
    \IF{$P_{(u,v)} > 0.5$}
	\STATE append $(u,v)$ to $E_{DG_m}$ with a weight $P_{(u,v)}$
    \ELSE
	\STATE append $(v,u)$ to $E_{DG_m}$ with a weight $1-P_{(u,v)}$
      \ENDIF
   \ENDFOR
\end{algorithmic}

In the next section, we analyze $DG$ to obtain final predicted order of arrival of nodes in $V_{G_m}$.
\subsection{Transformation of Directed Graph and Node Binning}

\hspace{0.17in} In this section, we process $DG$ obtained from the previous section to deduce the final prediction of order of arrival of nodes in $G_m$. Ideally we expect $DG$ to be acyclic in nature, as cycles would give rise to inconsistent prediction order among the nodes involved in the cycle. 
For example, lets say, $(u,v)$ and $(v,w)$ are in $E_{DG}$. This implies that $u$ has arrived before $v$ and $v$ has arrived before $w$. Hence, $w$ must have arrived before $u$. If $(w,u)$ also an edge, then it leads to a contradiction in the chronological ordering of $u$, $v$ and $w$. 
Since there is a fair possibility that $DG$ can be a cyclic graph, we intend to transform it into a Directed Acyclic Graph (DAG) and remove the inconsistencies involved. In our algorithm, we use a greedy technique to achieve the above.\\

The algorithm to transform $DG$ into $DAG$ is presented \\
below:
\begin{algorithmic}
\STATE Input: Directed Graph $DG$.
\STATE Output: Directed Acyclic Graph $DAG$.
\WHILE {$DG$  contains cycles}
  \STATE Remove the edge $(u,v)$ with the least $P_{(u,v)}$ : $(u,v) \in E_{DG}$. 
\ENDWHILE
\end{algorithmic}

The DAG thus obtained is free from inconsistencies. \\
$InDegree(v)$ represents the number of nodes that have been predicted to arrive after the arrival of $v$. Ideally, the node that had arrived earliest should have zero InDegree. The next earliest node should have an InDegree equal to 1 and so on. Since we are probabilistically simulating the growth environment of $G_m$, it is practically not always possible for the nodes to have the same sequence of InDegree as that of their order of arrival.

As the last step of the algorithm, we carry out the node binning process.
We find all the vertices $v \in V_{DAG}$ having the least $InDegree(v)$ and group them into a bin $B_1$. 
The binned vertices are then removed from $DAG$. 
We then repeat this step iteratively until there are no nodes left in $DAG$.
At each each iterative step $i$, we bin the nodes into a bin $B_i$.
By the ordering the bins according to their indices, we get the final predicted bin ordering.
    
Algorithm to bin the nodes from $DAG$ is presented below:

\begin{algorithmic}
\STATE Input: Directed Acyclic Graph $DAG$
\STATE Output: Bin Ordering
\STATE $count \leftarrow 1$ 
\WHILE {$|V_{DAG}| \neq 0$}
  	\STATE $minInDeg \leftarrow arg\ min (InDegree(u))\ where\ u \in V_{DAG}$
	\STATE Let $B_{count} \leftarrow \{u : \forall u \in V_{DAG}$\\ 
\hspace{1.2in}$and\ InDegree(u) = minInDeg\}$
	\STATE Remove all the nodes in $B_{count}$ from $V_{DAG}$\\
\hspace{1.2in}i.e, $V_{DAG} \leftarrow V_{DAG} - B_{count}$
	\STATE $Count \leftarrow Count + 1$ 
\ENDWHILE
\STATE Let $binOrdering \leftarrow [ B_1, B_2, B_3,...B_{Count}]$
\end{algorithmic}

$binOrdering$ gives the predicted chronological sequence of bins. The order of arrival of nodes within a bin is unknown. But the order of arrival of nodes across several such bins can be determined. The accuracy of this prediction, in contrast with accuracy of prediction using centrality measures, is discussed in the next section.

\section{Results and Discussions}
\subsection{Comparison between the predictions from Differential Core Ranking and \\ Plain Centrality}
\hspace{0.17in} Centrality Index of a vertex in a network indicates its relative importance in the network (refer section 2.4). Let $\chi$ be a base centrality measure. We hypothesize that, higher the relative importance of a vertex in a network $G_m$, earlier it has arrived during its evolution. Hence, the vertices in the network are arranged in the descending order of their $\chi$ centrality indices. Let this ordering of the nodes be denoted by $Plain\chi_{G_m}$. We apply DCR (refer section 4), with the same centrality $\chi$ as the base centrality, to the network $G_m$. The vertices in the network are arranged in the descending order of their DCR values. Let this ordering of the nodes be denoted by $Differential\chi_{G_m}$.

For experimental purposes, the actual order of arrival of nodes in $G_m$ is recorded. It is denoted by $list_{true}$. Let the predicted order be denoted by $list_{pred}$. To compute the accuracy of our prediction, we define a new quality measure called $\eta(list_{true}, list_{pred})$.

\begin{equation}
\eta(list_{true}, list_{pred}) = \frac{ n_c }{^{|V_{G_m}|}C_2}
\end{equation}

where $n_c$ is the number of pairs in $list_{pred}$ that are in correct relative order with respect to $list_{true}$.
To compare the prediction accuracy for the lists $Plain\chi_{G_m}$ and $Differential\chi_{G_m}$, we just compare the values of $\eta(list_{true}, Plain\chi_{G_m})$ and $\eta(list_{true}, Differential\chi_{G_m})$. In our experiments we consider the cases where $\chi$ represents Degree Centrality, Betweenness Centrality and Eigenvector Centrality.
The following figures represent the plots used to compare the values of $\eta(list_{true}, Plain\chi_{G_m})$ and $\eta(list_{true}, Differential\chi_{G_m})$ for varying number of nodes. Note that the number of connections $C_m$ is kept constant.

\begin{figure}[htp]
\centering
\includegraphics[scale=0.15]{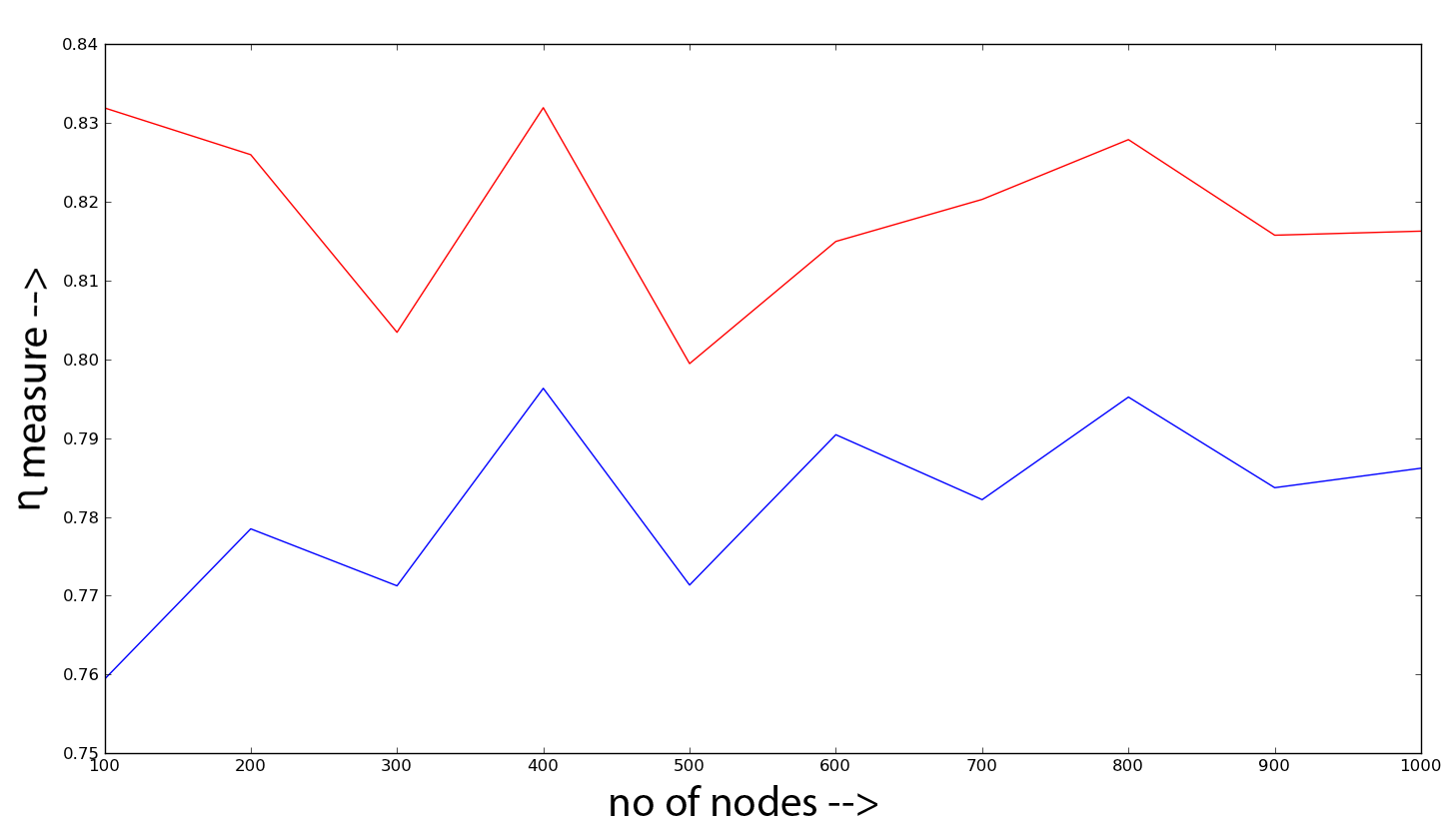}
\caption{Comparison of Differential Core Ranking (Red line), with
Betweenness as the base centrality measure, and Plain Betweenness
Centrality (Blue line) for the BA Networks with 3 connections. The
x-axis represents the number of nodes. The y-axis denotes
$\eta(list_{true}, DifferentialBetweenness_{G_m})$ and
$\eta(list_{true}, PlainBetweenness_{G_m})$.}
\label{}
\end{figure}

\begin{figure}[htp]
\centering
\includegraphics[scale=0.15]{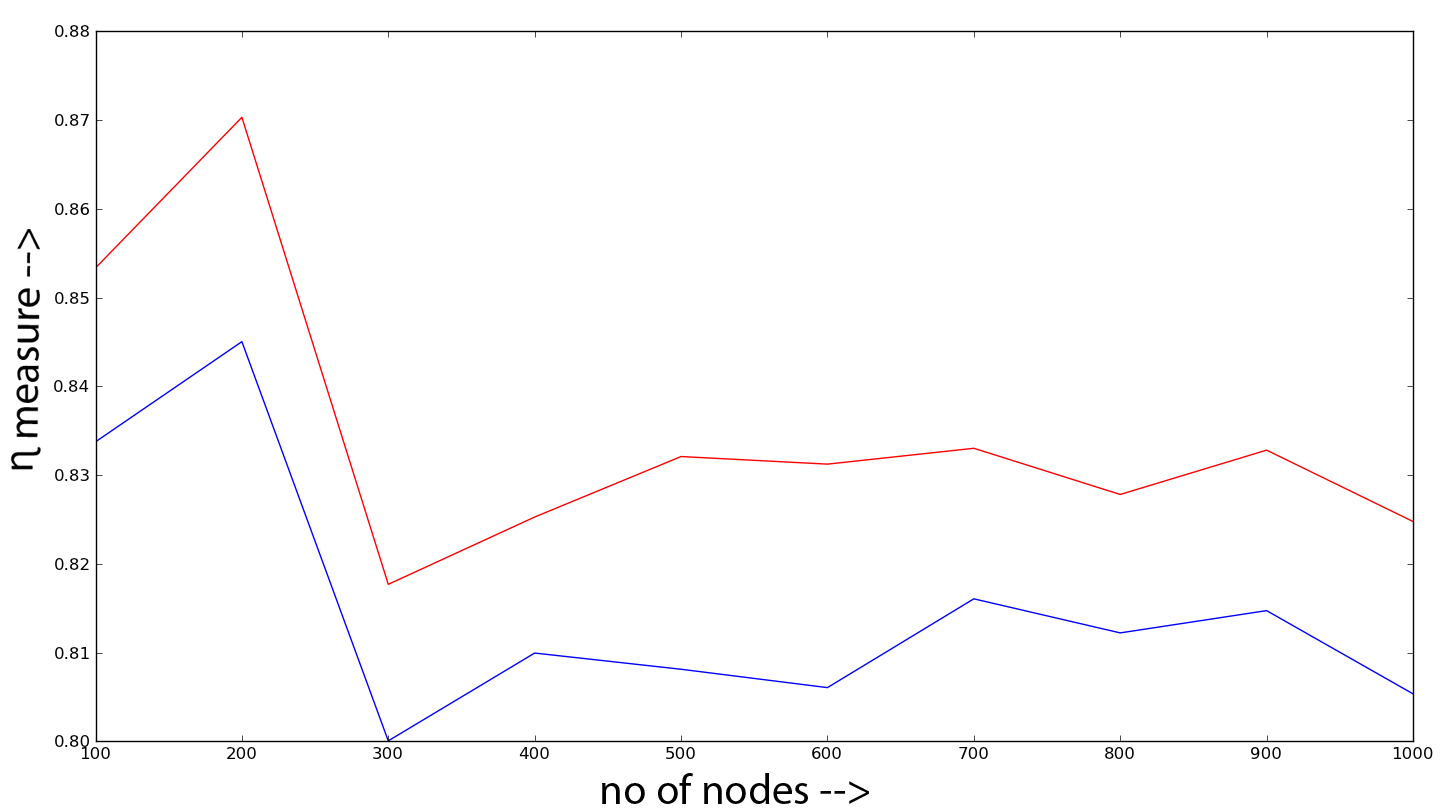}
\caption{Comparison of Differential Core Ranking (Red line), with
Degree as the base centrality measure, and Plain Degree Centrality
(Blue line) for the BA Networks with 3 connections. The x-axis
represents the number of nodes. The y-axis denotes $\eta(list_{true},
DifferentialDegree_{G_m})$ and $\eta(list_{true},
PlainDegree_{G_m})$.}
\label{}
\end{figure}

\begin{figure}[htp]
\centering
\includegraphics[scale=0.15]{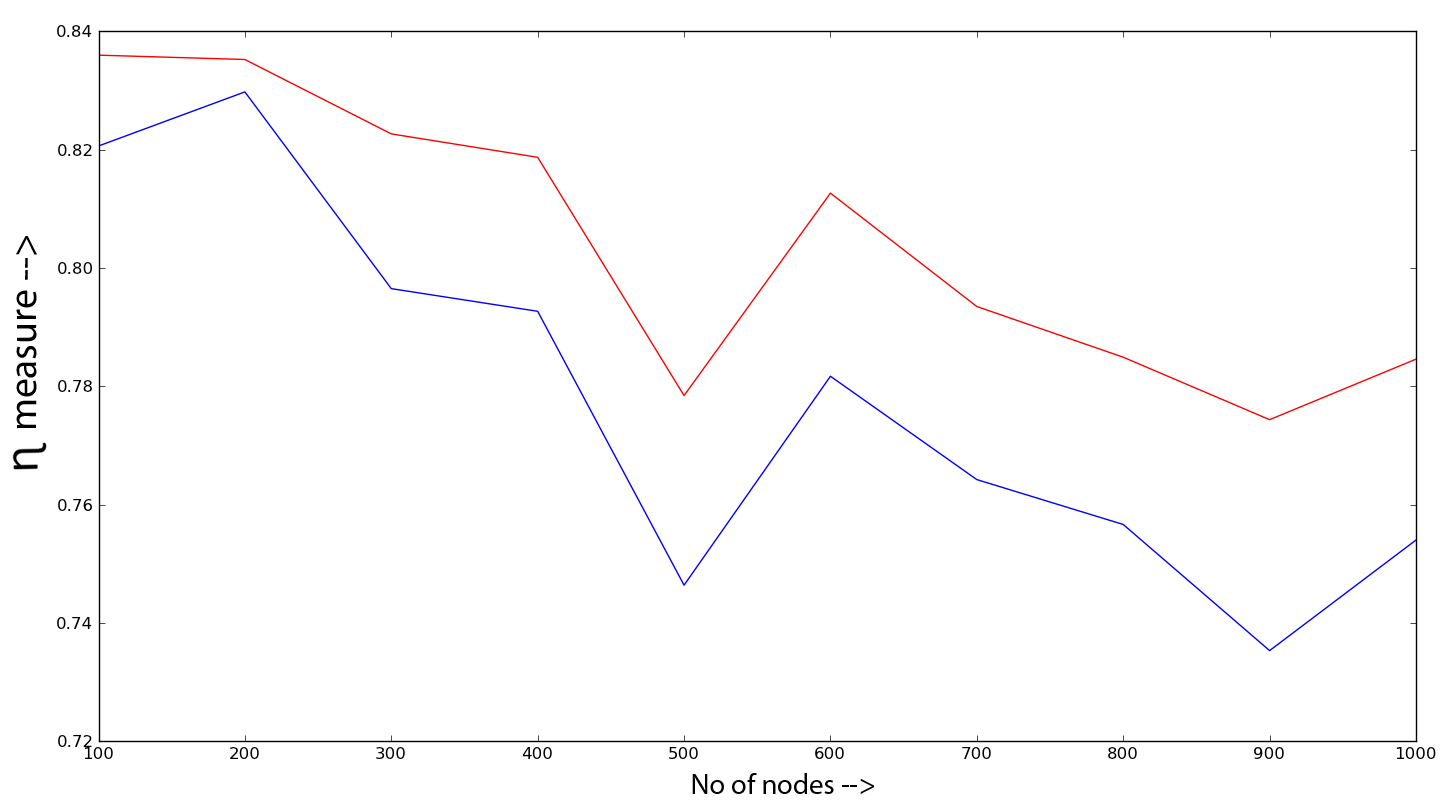}
\caption{ Comparison of Differential Core Ranking (Red line), with
Eigenvector as the base centrality measure, and Plain Eigenvector
Centrality (Blue line) for the BA Networks with 3 connections. The
x-axis represents the number of nodes. The y-axis denotes
$\eta(list_{true}, DifferentialEigen_{G_m})$ and $\eta(list_{true},
PlainEigen_{G_m})$.}
\label{}
\end{figure}

\begin{figure}[htp]
\centering
\includegraphics[scale=0.15]{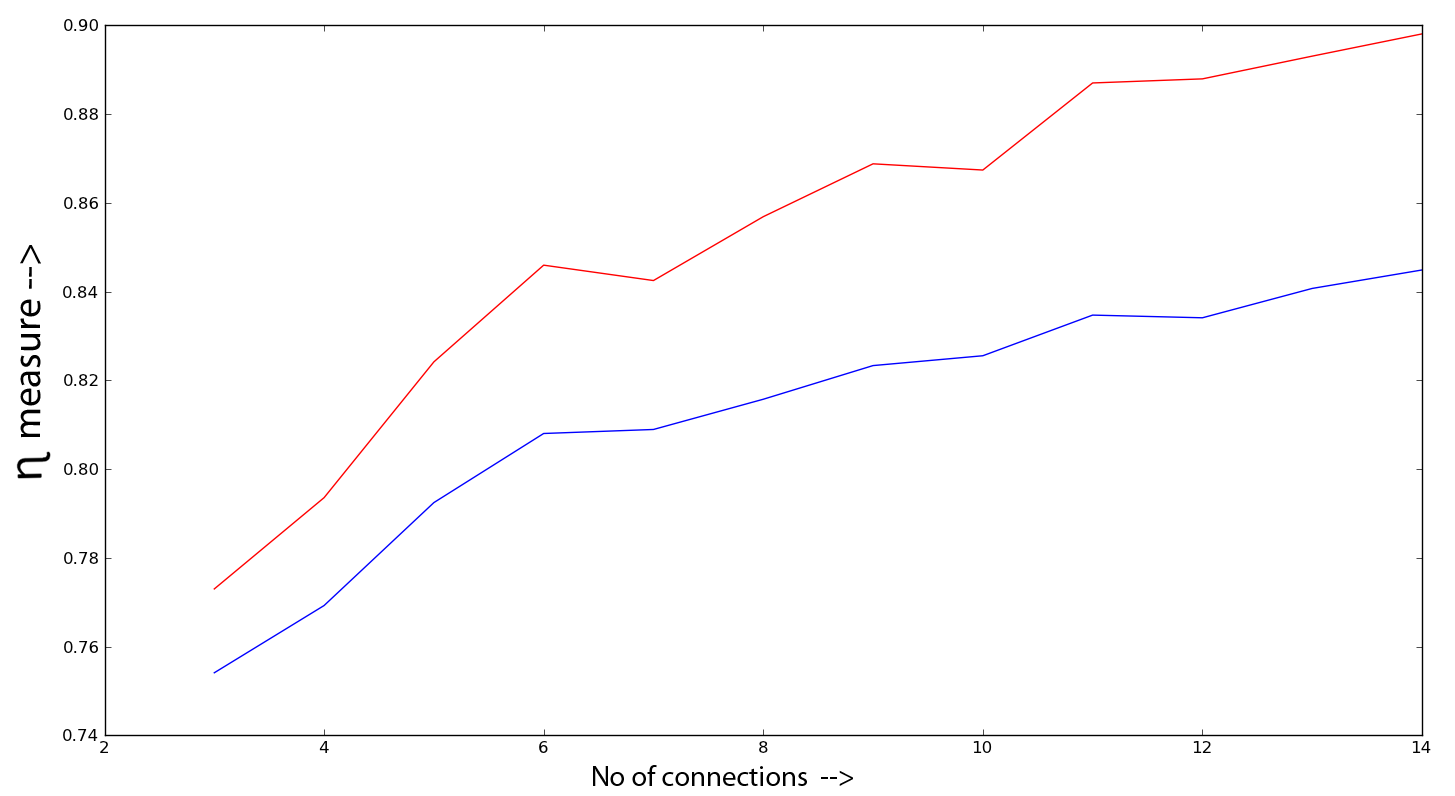}
\caption{Comparing Differential Core Ranking (Red line), with
betweenness as base the centrality measure, and Plain Betweenness
Centrality (Blue line) for the BA Networks with 1000 nodes. The x-axis
represents the connections $C_m$. The y-axis denotes
$\eta(list_{true}, DifferentialBetweenness_{G_m})$ and
$\eta(list_{true}, PlainBetweenness_{G_m})$.}
\label{}
\end{figure}

\begin{figure}[htp]
\centering
\includegraphics[scale=0.15]{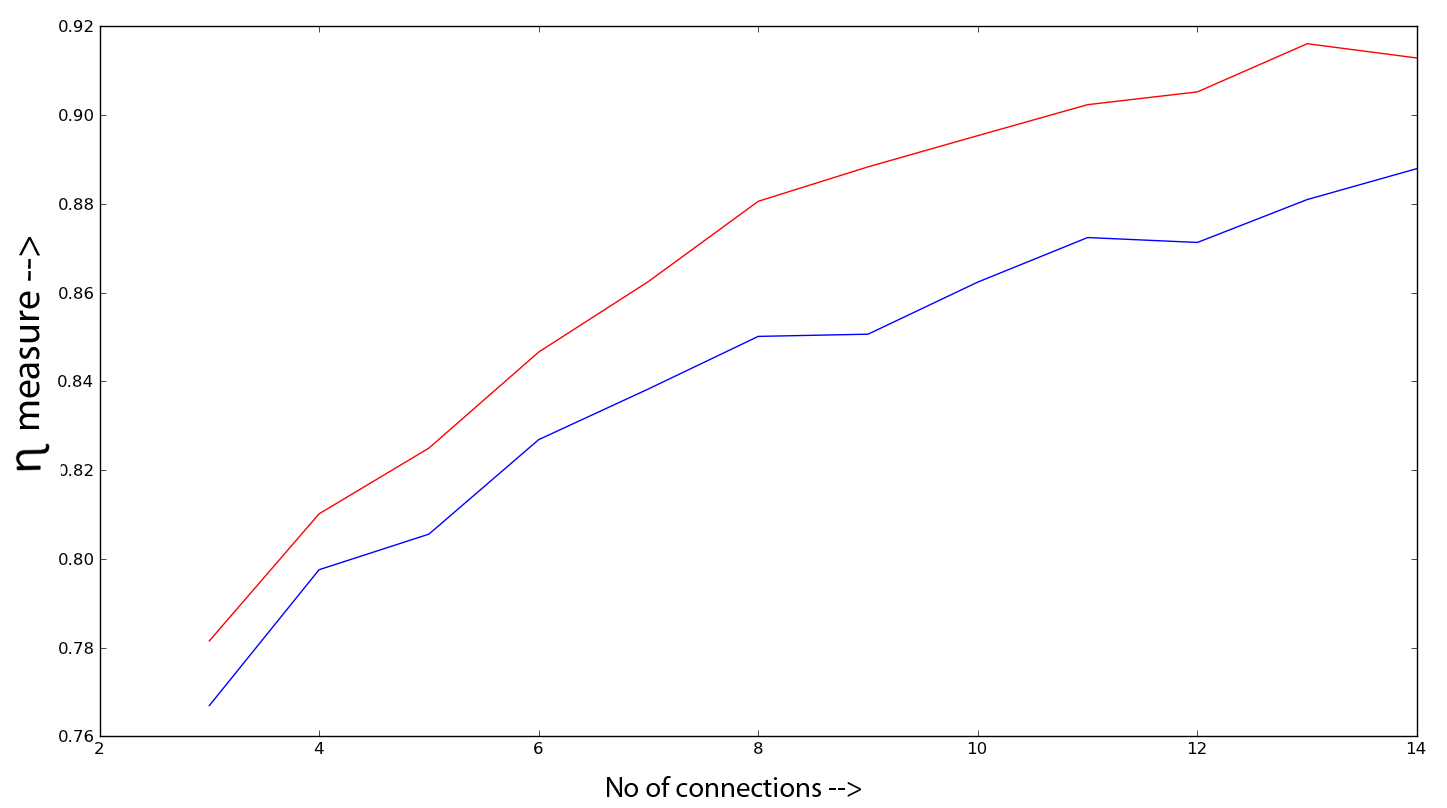}
\caption{ Comparison of Differential Core Ranking (Red line), with
Degree as the base centrality measure, and Plain Degree Centrality
(Blue line) for the BA Networks with 1000 nodes. The x-axis represents
the connections $C_m$. The y-axis denotes $\eta(list_{true},
DifferentialDegree_{G_m})$ and $\eta(list_{true},
PlainDegree_{G_m})$.}
\label{}
\end{figure}

\begin{figure}[htp]
\centering
\includegraphics[scale=0.15]{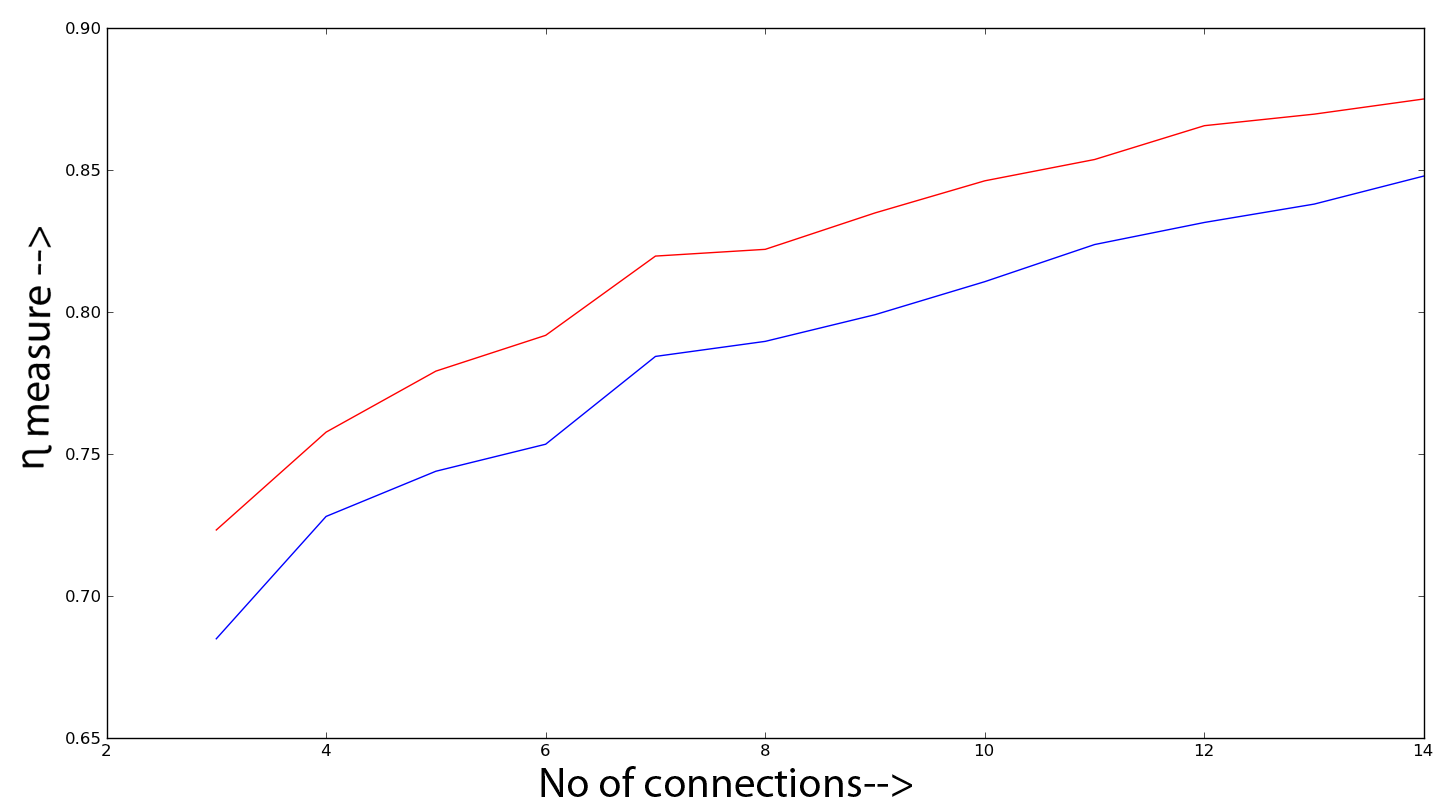}
\caption{Comparison of Differential Core Ranking (Red line), with
Eigenvector as the base centrality measure, and Plain Eigenvector
Centrality (Blue line) for the BA Networks with 1000 nodes. The x-axis
represents the connections $C_m$. The y-axis denotes
$\eta(list_{true}, DifferentialEigen_{G_m})$ and $\eta(list_{true},
PlainEigen_{G_m})$.}
\label{}
\end{figure}
\newpage
Figures 11 - 13 illustrate the performance of our alogrithm in
comparison with the centrality based binning, for varying number of
nodes in $G_m$. Figures 14-16 illustrate the performance of our
algorithm in comparision with centrality based binning, for varying
connection $C_m$ in $G_m$.
\subsection{Prediction of arrival order in every node pair with an attached probability} 

\hspace{0.17in} The outcome of section 5.3 is a weighted directed graph $DG$. We have associated a probability $P_{(u,v)}$ with every directed edge $(u,v) \in E_G$. $P_{(u,v)}$ indicates the probability with which $u$ has arrived before $v$. From the construction mechanism of $DG$, it is clear that $P_{(u,v)} > 0.5$. Closer the value of $P_{(u,v)}$ to 0.5, harder it is to ascertain the chronological ordering of $u$ and $v$. Note that there is a fair possibility that $DG$ can contain cycles. We claim that the inconsistencies in the prediction might be caused due to edges with $P_{(u,v)}$ close to 0.5. This may lead to a formation of cycles. 

We now present the analytical results that we have obtained, considering $G_m$ as reference network. We have generated $G_m$ using a BA model with 1000 nodes and 3 connections. We generate 50 synthetic networks. So, we set $\alpha = 50$. The analytical results thus obtained is given below:

\begin{figure}[htp]
\centering
\includegraphics[scale=0.35]{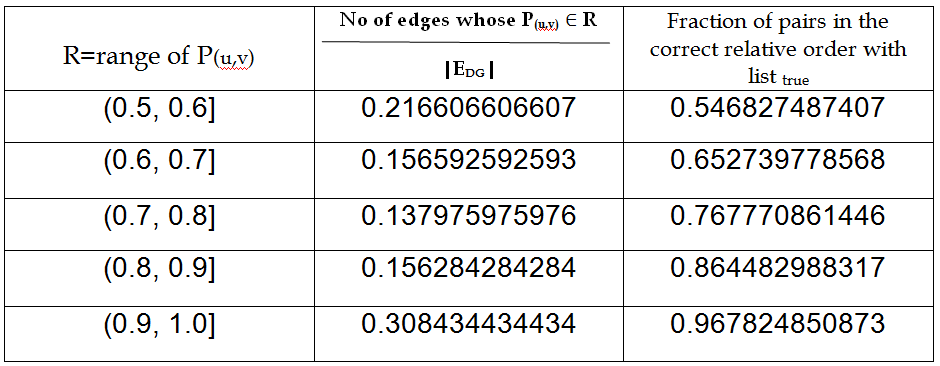}
\label{}
\end{figure}

Statistically, from the above table, we observe that the edges $(u,v)$ having $P_{(u,v)}$ in $(0.5, 0.6]$ constitute around 20\% of the edges. We also note that only around 50\% of these edges are in the correct relative order with $list_{true}$. Since a large fraction of edges belonging to this range are in incorrect relative ordering, they contribute to the cycle formation. Cycles introduce inconsistencies in node arrival order, hence they have to be removed. From our experiments, we have found out that $DG$ will become acyclic when we remove the edges $(u,v)$ continually in the increasing order until $P_{(u,v)} \approx 0.6$. We implement the same technique in section 5.4 to transform $DG$ to $DAG$.\\

Based on the facts and figures from the table, we observe that the fraction of pairs that are in correct relative order with $list_{true}$ increases as the sampled range increases. Hence we conclude that, higher $P_{(u,v)}$ implies a stronger notion of relative ordering of $(u,v)$.

\subsection{Comparison between the predictions from DCR binning and Plain Centrality binning}
\hspace{0.17in} The end result of our method (section 5.4) is the ordering of the bins, referred to as $binOrdering_{DCR\chi}$. Let $\Delta$ be the number of bins in $binOrdering_{DCR\chi}$. Let $\eta_{DCR\chi}$ denote the BQM value of $binOrdering_{DCR\chi}$, where $\chi$ refers to the base centrality measure for DCR. 

We derive the $binOrdering_\chi$ (refer section 3.2) with $\Delta$ number of bins, and $\chi$ indicating the centrality measures.  Let $binOrdering_{betweenness}$, $binOrdering_{eigen}$ and  $binOrdering_{degree}$ denote the chronology of bins with $\chi$ set as Betweenness, Eigenvector and Degree Centralities respectively.

Let $\eta_{betweenness}$, $\eta_{eigen}$ and $\eta_{degree}$ denote the BQM value of $binOrdering_{betweenness}$, $binOrdering_{eigen}$ and  $binOrdering_{degree}$ respectively. Finally, we compare $\eta_{betweenness}$, $\eta_{eigen}$, $\eta_{degree}$ and $\eta_{DCR\chi}$ where $\chi$ is the base centrality (refer section 4).

\begin{figure}[htp]
\centering
\includegraphics[scale=0.16]{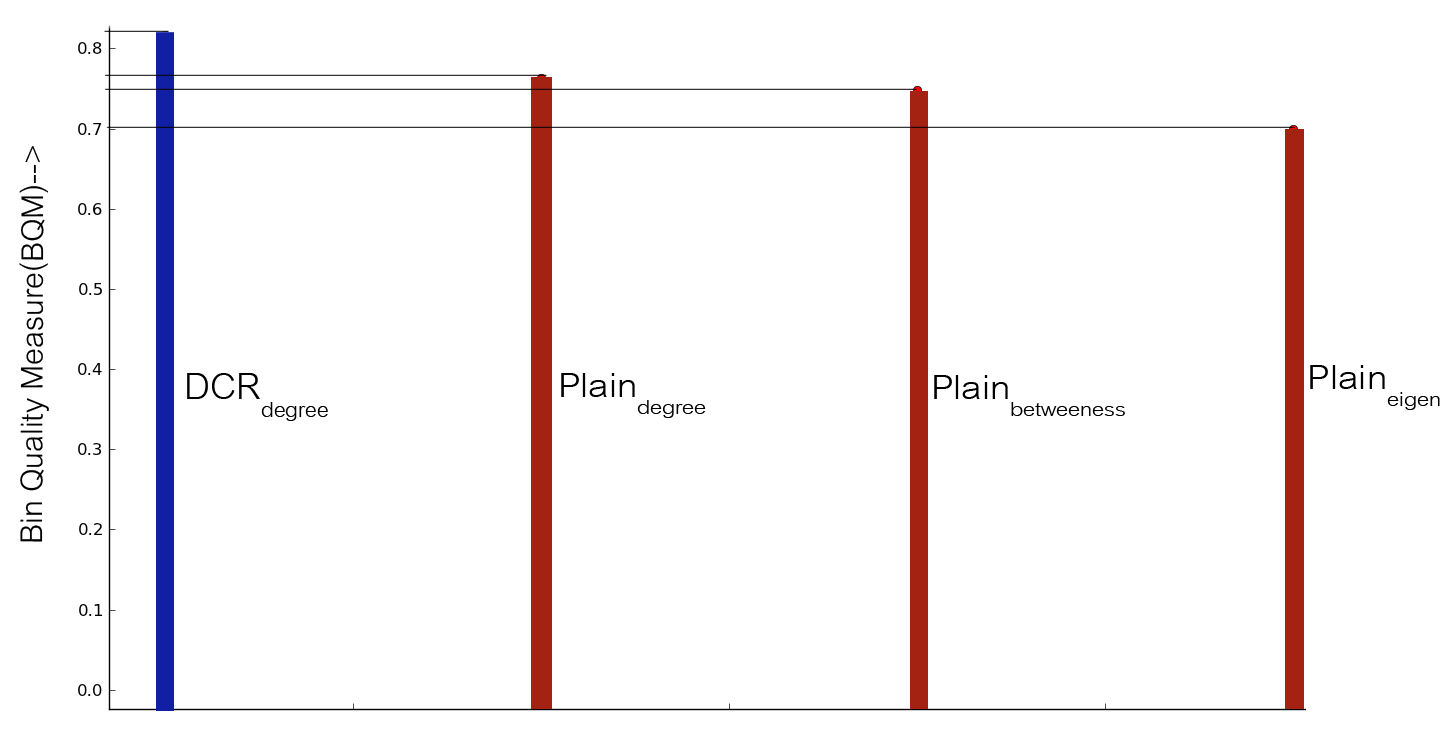}
\caption{The plot denotes the BQM score for various binning methodologies for the reference graph $G_m$ of 1000 nodes and 3 connections. In our experiment, we have set $\alpha = 50$.The results we obtained are as follows:
$\eta_{DCR_{degree}}=0.804513946531$
$\eta_{degree}=0.767615011251$
$\eta_{betweenness}=0.759827243464$
$\eta_{eigen}=0.695466553648$ 
number of bins=91}
\label{}
\end{figure}

\begin{figure}[htp]
\centering
\includegraphics[scale=0.16]{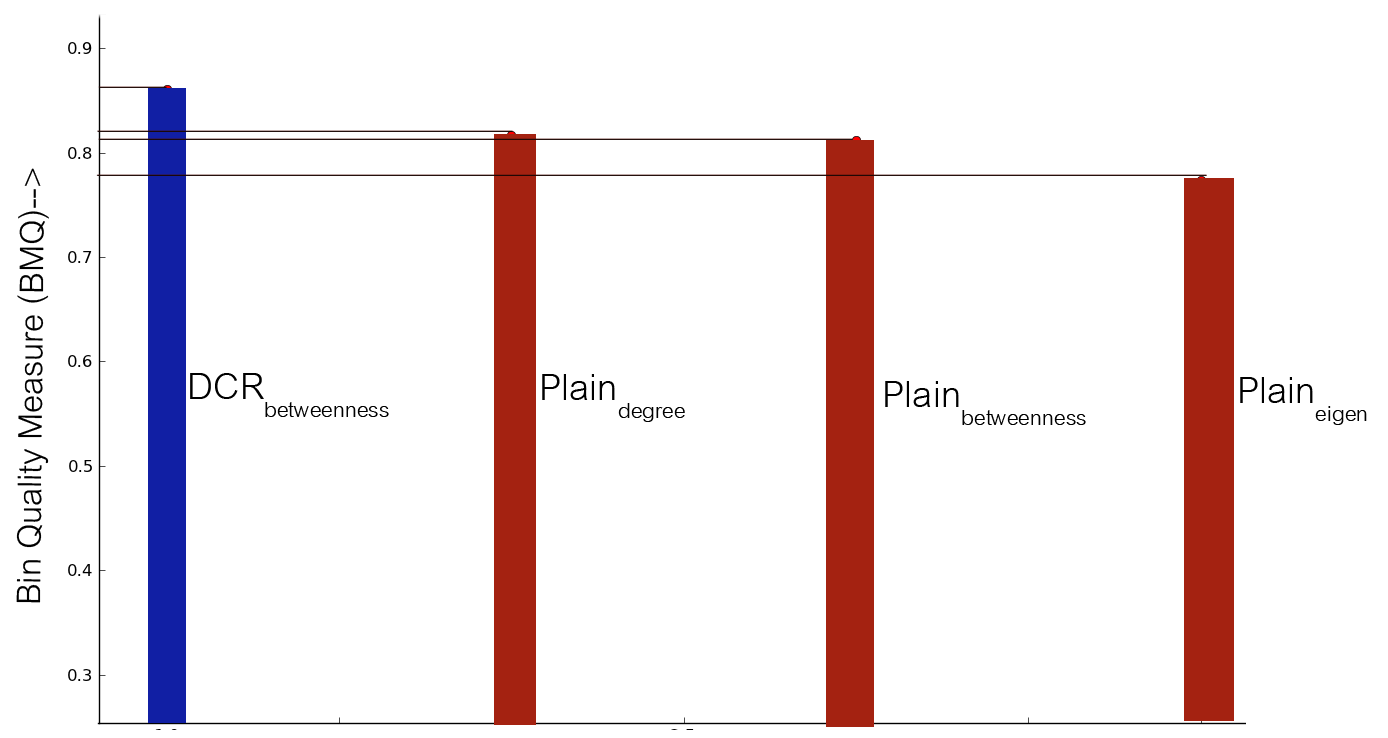}
\caption{The plot denotes the BQM score for various binning methodologies for the reference graph $G_m$ of 1000 nodes and 3 connections. In our experiment, we have set $\alpha = 50$.The results we obtained are as follows:
$\eta_{DCR_{betweenness}}=0.87153926121$
$\eta_{degree}=0.8251012352$
$\eta_{betweenness}=0.8158246115$
$\eta_{eigen}=0.7823167778$ 
number of bins=63}
\label{}
\end{figure}

\begin{figure}[htp]
\centering
\includegraphics[scale=0.16]{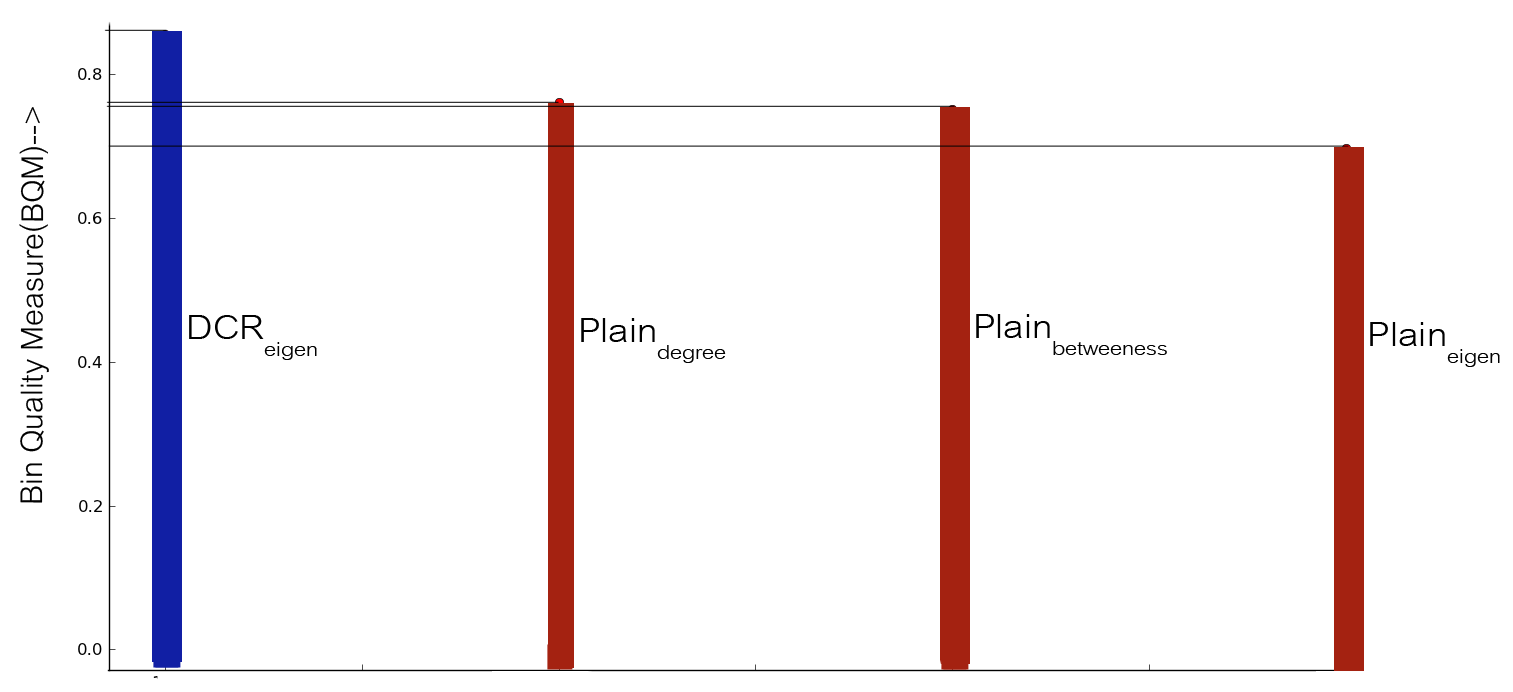}
\caption{The plot denotes the BQM score for various binning methodologies for the reference graph $G_m$ of 1000 nodes and 3 connections. In our experiment, we have set $\alpha = 50$.The results we obtained are as follows:
$\eta_{DCR_{eigen}}=0.84654821986$
$\eta_{degree}=0.7697124538121$
$\eta_{betweenness}=0.753169421166$
$\eta_{eigen}=6899122714632$ 
number of bins=77}
\label{}
\end{figure}

\newpage
Predicted chronological sequence of bins obtained from $DCR$ For any $\chi$ as base centrality, we observe that it is more accurate compared to any other centrality based approaches.

\section{Conclusions}
We presented a novel framework for uncovering the precursor of a SFN evolved by preferential attachment model. Our approach involves synthesis of many such SFNs, mapping these SFNs with the reference network based on $DCR$ score associated with the nodes and arriving at the final predicted order. We presented results based on a novel indexing method called the differential core ranking, which proved to provide better node arrival prediction than the approaches based on standard centrality measures. 

Our approach can be put to practice in situations where one is given a real world network (which is known to have evolved by preferential attachment) and one is interested to obtain the order of node arrivals. A useful application would be to unravel the age of the links in www network, which is known to be scale-free~\cite{barabasi02} . Also, knowing the age of the nodes in a disease spreading network would help us determine the susceptibility of the nodes to get infected. For example, a newly arrived node is more susceptible to be infected as opposed to a node that has been present in the network for long. Such a node might have possibly developed the necessary immunity to counter the infection. Our results show that, if a network is known to have evolved in steps, then its chronology can be effectively excavated.  
\bibliographystyle{plain}
\bibliography{prediction}

\begin{thebibliography}{10}

\bibitem{barabasi02}
R.~Albert and A.~L. Barab\'{a}si.
\newblock {Statistical mechanics of complex networks}.
\newblock {\em Reviews of Modern Physics}, 74(1):47--97, January 2002.

\bibitem{anthonisse71}
J.M. Anthonisse.
\newblock The rush in a directed graph.
\newblock Technical Report BN 9/71, Stichting Mathematisch Centrum, 1971.

\bibitem{barabasi99-1}
A.~L. Barab\'{a}si and R.~Albert.
\newblock {Emergence of Scaling in Random Networks}.
\newblock {\em Science}, 286(5439):509--512, October 1999.

\bibitem{bavelas50}
A.~Bavelas.
\newblock {Communication Patterns in Task-Oriented Groups}.
\newblock {\em The Journal of the Acoustical Society of America},
  22(6):725--730, 1950.

\bibitem{bonacich72}
P.~Bonacich.
\newblock {Factoring and weighting approaches to status scores and clique
  identification}.
\newblock {\em Journal of Mathematical Sociology}, 2(1):113--120, 1972.

\bibitem{borgatti06}
S.~Borgatti and M.~Everett.
\newblock A graph-theoretic perspective on centrality.
\newblock {\em Social Networks}, 28(4):466--484, October 2006.

\bibitem{brandes01}
U.~Brandes.
\newblock A faster algorithm for betweenness centrality.
\newblock {\em Journal of Mathematical Sociology}, 25:163--177, 2001.

\bibitem{zweig05}
U.~Brandes and T.~Erlebach.
\newblock {Network Analysis. Methodological Foundations}.
\newblock {\em Network Analysis, Lecture Notes in Computer Science}, 3418,
  2005.

\bibitem{freeman77}
L.~C. Freeman.
\newblock A set of measures of centrality based on betweenness.
\newblock {\em Sociometry}, 40(1):35--41, 1977.

\bibitem{sanket11}
Saket Navlakha and Carl Kingsford.
\newblock {Network Archaeology: Uncovering Ancient Networks from Present-Day
  Interactions}.
\newblock {\em PLoS Comput Biol}, 7(4):e1001119+, April 2011.

\end{thebibliography}
\end{document}